\documentclass[twocolumn,english,nofootinbib]{revtex4-1}
\usepackage[T1]{fontenc}
\usepackage[latin9]{inputenc}
\usepackage[a4paper]{geometry}
\usepackage{amsmath}
\usepackage{amssymb}
\usepackage{esint}
\usepackage{hyperref}
\usepackage{babel}
\makeatletter

\allowdisplaybreaks

\makeatother

\begin{document}
\title{Massive strings from a haunted field theory}
\author{Nicholas Carabine}
\author{Renann Lipinski Jusinskas}
\affiliation{FZU - Institute of Physics of the Czech Academy of Sciences \& CEICO ~\\
Na Slovance 2, 18221 Prague, Czech Republic}

\begin{abstract}
In this work we present the $\alpha'$-exact background equations of motion of the bosonic chiral string (also known as Hohm-Siegel-Zwiebach model), with the spin two ghost fields integrated out. This is the first instance of a worldsheet model in which all corrections are fully determined in a generic curved spacetime. As a concrete cross-check, we find complete agreement between all three-point and a sample of four-point tree level scattering amplitudes computed using field theory methods and the chiral string prescription. These equations of motion provide a field theoretical shortcut to compute worldsheet correlators in conventional bosonic strings (with arbitrary number of massless and mass level one states), and outline a new perspective on massive resonances in string theory.\end{abstract}
\maketitle

\section{Overview}

Many deep and unexpected connections in physics were uncovered due
to the fruitful, though not often evident, interchange between strings and fields.

An explicit example is the fact that some scattering
amplitudes in quantum field theory have a dual description in terms
of integrations over two-dimensional (Riemann) surfaces with marked
points. This is the basic idea underlying the Cachazo-He-Yuan amplitudes
(CHY) \cite{Cachazo:2013hca}, flagrantly reminiscent of string theory
and its worldsheet. However, conventional string theory has
an infinite physical spectrum and cannot be described through ordinary
field theory methods. There is no finite Lagrangian for such an enterprise.

When focusing on massless theories, CHY amplitudes admit a more
fundamental, string-like description in terms of ambitwistor strings
\cite{Mason:2013sva}. In spite of living in two dimensions,
ambitwistor strings are chiral theories. Their constituents are
function of only one of the light-cone directions of the worldsheet.
In addition, ambitwistor strings have no dimensionful parameter, hence
a massless spectrum. Indeed, they are related to a chiral model
proposed by Hohm, Siegel, and Zwiebach (HSZ) \cite{Hohm:2013jaa}
through a tensionless limit \cite{Siegel:2015axg,Casali:2016atr,Azevedo:2017yjy,Kalyanapuram:2021xow,Kalyanapuram:2021vjt}.

The HSZ model is supposed to be the worldsheet theory underlying an effective spacetime description of the massless string spectrum, with manifest T-duality.  It was
argued to be a consistent truncation of string theory with complete control over the corrections in $\alpha'$ (the string length squared) \cite{Hohm:2013jaa,Hohm:2014xsa,Hohm:2016lim}. As shown in \cite{Siegel:2015axg} (see
also \cite{LipinskiJusinskas:2019cej} for further details), this
model may be derived from the first order Polyakov action
through a singular gauge choice, with a degenerate worldsheet metric. In this sense, chiral strings are more than a particle
but less than a string. Though containing ghosts, their physical spectrum
is finite, suggesting a possible Lagrangian description.

The heterotic HSZ model has been recently used to determine 
conformal field theory (CFT) correlators in conventional
string theory amplitudes \cite{Guillen:2021mwp}. Those were tree level results, involving
an arbitrary number of massless fields and only one state from the
first massive level. More intriguingly, they were derived using a mix
of string theory techniques and a field theory involving ghosts. In
hindsight, the appearance of ghosts should not come as a surprise.
The impressive consistency of string theory relies on an infinite
tower of massive states. But when trying to describe specific mass
levels through a Lagrangian the emergence of a ``sick'' field theory
may be expected.

The breakthrough of \cite{Guillen:2021mwp} was at the time limited
by two factors. First, a lack of control of the gravitational sector of the theory. Second, the unique massive multiplet of the chiral string,  matching the degrees of freedom of the first mass level of the open superstring. While
higher mass levels may now be individually accessed through
the asymetrically twisted strings \cite{Jusinskas:2021bdj}, their
field theory realization is more involved, naturally containing higher
spins. The bosonic chiral string, on the other hand, offers the possibility
of the \emph{exact} computation of its tree level dynamics. That includes
the usual massless bosonic string spectrum (graviton, Kalb-Ramond
two-form, dilaton) and two spin-two massive states -- the ghosts of
the HSZ model. This possibility stems from the fact that the worldsheet
action is free while the coupling to the background is exclusively
governed by the BRST charge. Its nilpotency should yield the field equations describing the spacetime dynamics of the physical spectrum.

In this work we present for the first time the exact equations of
motion of the HSZ model (bosonic chiral string). The respective field
theory has a six derivative kinetic term after integrating out the
auxiliary spin-two fields, including Chern-Simons-like corrections
for the Kalb-Ramond field-strength \cite{Green:1984sg,Hohm:2014eba} and the complete expressions
for the $\alpha'$ corrections. In analogy with the results of \cite{Guillen:2021mwp},
we provide a field theoretical shortcut to obtain the relevant CFT
correlators for computing $N$-point tree level (open and closed)
bosonic string amplitudes with an arbitrary number of massless and
mass level one states. This is a concrete step to understand the role of mass in string theory from a novel perspective. One in which mass levels are singled out, with a dual description as ghosts in a gravitational theory.

\section{Bosonic chiral strings}

We start with an alternative description of the HSZ model. It can
be viewed as an ambitwistor string in which one of the states of
the zero momentum cohomology acquires a vacuum expectation value.
In order to see this, consider the ambitwistor gauge fixed action
and energy-momentum tensor \cite{Mason:2013sva},
\begin{eqnarray}
S_{0} & = & \tfrac{1}{2\pi}\int dz d\bar{z}(P_{m}\bar{\partial}X^{m}+b\bar{\partial}c+\tilde{b}\bar{\partial}\tilde{c}),\\
T & = & -P_{m}\partial X^{m}-b\partial c-\partial(bc)-\tilde{b}\partial\tilde{c}-\partial(\tilde{b}\tilde{c}),
\end{eqnarray}
where $\partial\equiv\partial/\partial z$, $\bar{\partial}\equiv\partial/\partial \bar{z}$. The target space coordinates
are denoted by $X^{m}$, with canonical conjugate $P_{m}$. The ghost
pairs $\{b,c\}$ and $\{\tilde{b},\tilde{c}\}$ are respectively associated
with the generators $T$ and $H=-\tfrac{1}{2}\eta^{mn}P_{m}P_{n}$,
where $\eta_{mn}$ is the flat space metric. The BRST charge is simply
$Q_{0}=\oint(cT-bc\partial c+\tilde{c}H)$, with $\{Q_0,b\}=T$ and $\{Q_0,\tilde{b}\}=H$. Requiring its nilpotency
fixes the spacetime dimension to be $d=26$. 

Physical states are defined to be in the ghost number two cohomology
of $Q_{0}$, annihilated by the zero mode of the $b$ ghost \cite{Berkovits:2018jvm}.
In particular, there are two zero-momentum states, 
\begin{eqnarray}
U_{h} & = & c\tilde{c} h^{mn} P_{m}P_{n},\\
U_{t} & = & c\tilde{c}\eta_{mn}\partial X^{m}\partial X^{n}-2bc\tilde{c}\partial\tilde{c}-3\tilde{c}\partial^{2}\tilde{c},
\end{eqnarray}
with relative dimension quartic in length. The first generates constant deformations of
the flat space metric $\eta^{mn} \to \eta^{mn} + h^{mn}$, as it couples to the BRST current
as $[b_{-1}, U_{h}]=\tilde{c}h^{mn}P_{m}P_{n}$. Similarly, the coupling of
$U_{t}$ is given by
\begin{equation}
[b_{-1}, U_{t}]=\tilde{c}(\eta_{mn}\partial X^{m}\partial X^{n}+2b\partial\tilde{c}).
\end{equation}
It generates a \textit{tensile} deformation that will be parametrized by $(\alpha')^{2}$, becoming the only dimensionful parameter of the model.  In this case, the new BRST charge may be  expressed as
\begin{equation}
Q=\oint(cT+\tilde{c}\mathcal{H}-bc\partial c+4c\tilde{b}\partial\tilde{c}+2c\partial\tilde{b}\tilde{c}),\label{eq:BRST-charge}
\end{equation}
where
\begin{multline}
\mathcal{H}=-\tfrac{1}{2}(P_{m}P_{n}\eta^{mn}+\tfrac{1}{(\alpha')^{2}}\partial X^{m}\partial X^{n}\eta_{mn})\\
-4\tilde{b}\partial c-2\partial\tilde{b}c-\tfrac{1}{(\alpha')^{2}}(b\partial\tilde{c}+\tfrac{1}{2}\partial b\tilde{c}).
\end{multline}

The gauge algebra of $T$ and $\mathcal{H}$ is given by\begin{subequations}\label{eq:TH-algebra}
\begin{eqnarray}
T(z)\,T(y) & \sim & \frac{2T}{(z-y)^{2}}+\frac{\partial T}{(z-y)},\label{eq:TT}\\
T(z)\,\mathcal{H}(y) & \sim & \frac{2\mathcal{H}}{(z-y)^{2}}+\frac{\partial\mathcal{H}}{(z-y)},\label{eq:TH}\\
\mathcal{H}(z)\,\mathcal{H}(y) & \sim & \tfrac{1}{(\alpha')^{2}}\bigg(\frac{2T}{(z-y)^{2}}+\frac{\partial T}{(z-y)}\bigg).\label{eq:HH}
\end{eqnarray}
\end{subequations}It can be seen as two copies of a chiral Virasoro
algebra with generators $T_{\pm}=(T\pm\alpha'\mathcal{H})/2$, and
gives rise to a sectorized interpretation of the chiral string \cite{LipinskiJusinskas:2019cej}.

The physical states of the chiral model have been discussed in detail in
\cite{Azevedo:2019zbn} (in the second order formalism, they can be nicely described with the usual oscillator construction \cite{Lee:2017utr,Lee:2017crr}). The massless level is the conventional bosonic
string spectrum (graviton, Kalb-Ramond two-form, dilaton), and there
are two spin 2 states with mass $m^{2}=\pm4/\alpha'$. The precise
form of the vertex operator $U$ is not immediately relevant, but
we are interested in the corresponding deformation of the BRST charge.
Up to BRST-exact contributions and total derivatives, it can be generically expressed as
\begin{multline}
[b_{-1}, U]=\tilde{c}(\partial^{2}X^{m}A_{m}+P_{m}\partial X^{n}B_{n}^{m}+\partial P_{m}C^{m})\\
-\tfrac{1}{2}\tilde{c}(P_{m}P_{n}g^{mn}+\tfrac{1}{(\alpha')^{2}}\partial X^{m}\partial X^{n}\tilde{g}_{mn})+c\partial^{2}D.\label{eq:BRST-deformation}
\end{multline}
The fields $\{A_{m},B_{n}^{m},C^{m},D,g^{mn},\tilde{g}_{mn}\}$ will
then describe the background dynamics.

\section{Curved Spacetime}

In the bosonic chiral string, the coupling to the curved background
is determined solely in terms of deformations of the BRST charge.
Motivated by (\ref{eq:BRST-deformation}), our idea is to propose
modified operators $T$ and $\mathcal{H}$ that still satisfy the
algebra (\ref{eq:TH-algebra}). This will ensure the nilpotency of
the BRST charge (\ref{eq:BRST-charge}) while imposing consistency
conditions -- the field equations -- for the background. At a semiclassical level, a similar procedure has been
long known in conventional (super) string theory \cite{Banks:1986fu,Maharana:1986ix,Das:1989wf,Berkovits:2001ue}.
More recently, it has been extended to ambitwistor strings in NS-NS
backgrounds \cite{Adamo:2014wea}.

The new operators are defined as
\begin{equation}
T=-P_{m}\partial X^{m}-b\partial c-\partial(bc)-\tilde{b}\partial\tilde{c}-\partial(\tilde{b}\tilde{c})+\partial^{2}D,\label{eq:EMoperator}
\end{equation}
\begin{multline}
\mathcal{H}=-\tfrac{1}{2}(P_{m}P_{n}g^{mn}+\tfrac{1}{(\alpha')^{2}}\partial X^{m}\partial X^{n}\tilde{g}_{mn})\\
+\partial^{2}X^{m}A_{m}+P_{m}\partial X^{n}B_{n}^{m}+\partial P_{m}C^{m}\\
-4\tilde{b}\partial c-2\partial\tilde{b}c-\tfrac{1}{(\alpha')^{2}}[b\partial\tilde{c}+\tfrac{1}{2}\partial b\tilde{c}].\label{eq:Hoperator}
\end{multline}
The spacetime metric is $g_{mn}$ (with inverse $g^{mn}$). For later convenience we decompose $B_{n}^{m}$
as
\begin{multline}
B_{n}^{m}=g_{np}S^{pm}+\tfrac{1}{\alpha'}b_{np}g^{pm}\\
+\tfrac{1}{4}g^{pm}(\partial_{p}g^{qr}\partial_{r}g_{nq}-\partial_{n}g^{qr}\partial_{r}g_{pq})\\
+\tfrac{1}{4}g^{pm}(\partial_{r}g^{qr}\partial_{n}g_{pq}-\partial_{r}g^{qr}\partial_{p}g_{nq})\\
+\tfrac{1}{2}g^{pm}(\partial_{n}g^{qr}g_{pq}\partial_{r}D-\partial_{p}g^{qr}g_{nq}\partial_{r}D),\label{eq:Bdecomposition}
\end{multline}
with $S^{mn}$ and $b_{mn}$ symmetric and antisymmetric in $m\leftrightarrow n$,
respectively and without loss of generality.

There are slightly more general deformations of $\mathcal{H}$ that spoil the $T-\mathcal{H}$ algebra but still lead to a nilpotent BRST charge. However, they correspond to pure gauge degrees of freedom in the field theory setup. Towards  \eqref{eq:TH-algebra}, first notice that the OPE \eqref{eq:TT} is left unchanged using \eqref{eq:EMoperator}. Next, we have to impose that the new $\mathcal{H}$ is a primary conformal operator, cf. equation \eqref{eq:TH}. A direct
computation leads to
\begin{multline}
T(z)\,\mathcal{H}(y)\sim\frac{2\mathcal{H}}{(z-y)^{2}}+\frac{\partial\mathcal{H}}{(z-y)}\\
+\tfrac{1}{2}\frac{(\partial X^{m}\alpha_{m}+P_{m}\chi^{m})}{(z-y)^{3}}-\frac{\beta}{(z-y)^{4}},
\end{multline}
with
\begin{eqnarray}
\alpha_{m} & = & A_{m}+B_{m}^{n}\partial_{n}D-\tfrac{1}{2}\partial_{n}B_{m}^{n}\nonumber \\
 &  & -\tfrac{1}{2}g^{np}\partial_{m}\partial_{n}\partial_{p}D+C^{n}\partial_{m}\partial_{n}D,
\\
\beta & = & B_{m}^{m}+3g^{mn}\partial_{m}\partial_{n}D +2\partial_{m}C^{m}\nonumber \\
 &  & -6C^{m}\partial_{m}D,\label{eq:THquartic}\\
\chi^{m} & = & C^{m}+\tfrac{1}{2}\partial_{n}g^{mn}-g^{mn}\partial_{n}D.
\end{eqnarray}
$\alpha_{m}=\chi^{m}=0$ provide algebraic solutions
for $A_{m}$ and $C^{m}$, while $\beta=0$ yields a scalar equation
of motion.

Finally, the OPE of $\mathcal{H}$ in (\ref{eq:Hoperator}) with itself,
c.f. (\ref{eq:HH}), can be computed to be
\begin{multline}
\mathcal{H}(z)\,\mathcal{H}(y)\sim\tfrac{1}{(\alpha')^{2}}\bigg(\frac{2T}{(z-y)^{2}}+\frac{\partial T}{(z-y)}\bigg)\\
+\frac{1}{(z-y)^{2}}\partial X^{m}\partial X^{n}\beta_{mn}^{G}+\frac{1}{(z-y)^{2}}P_{m}\partial X^{n}\beta_{n}^{m}\\
+\frac{2}{(z-y)^{2}}P_{m}P_{n}\Sigma^{mn}+\frac{\Sigma}{(z-y)^{4}}+\ldots.
\end{multline}
The ellipsis denote other poles in the OPE that are not independent
from $\{\alpha_{m},\beta,\beta_{n}^{m},\beta_{mn}^{G},\chi^{m},\Sigma,\Sigma^{mn}\}$.
For instance, there is a cubic pole with numerator $\partial\Sigma/2$.

The coefficient $\Sigma^{mn}$ can be expressed as
\begin{multline}
\Sigma^{mn}=S^{mn}-\tfrac{1}{2}(C^{p}\partial_{p}g^{mn}+\partial_{q}g^{mp}\partial_{p}g^{nq})\\
+\tfrac{1}{4}g^{pq}\partial_{p}\partial_{q}g^{mn}=0,\label{eq:PPeom}
\end{multline}
implying that $S^{mn}$ is solved
in terms of $g^{mn}$ and $C^{m}$. Then, after defining the dilaton field as
\begin{equation}
\phi\equiv D-\tfrac{1}{4}\ln g,\label{eq:dilaton-def}
\end{equation}
where $g=\det(g_{mn})$, equation (\ref{eq:THquartic}) becomes
\begin{equation}
5\Box\phi=6\nabla_{m}\phi\nabla^{m}\phi-R,\label{eq:dilaton-eom}
\end{equation}
Here $\Box\equiv g^{mn}\nabla_{m}\nabla_{n}$, with covariant derivative
$\nabla_{m}$. $\alpha'$ corrections to the dilaton equation
of motion appear through the scalar curvature $R$.

The coefficient $\beta_{n}^{m}$, given by\begin{widetext}
\vspace{-0.5cm}
\begin{multline}
\beta_{n}^{m}=\tfrac{2}{(\alpha')^{2}}(\delta_{n}^{m}-g^{mp}\tilde{g}_{np})+\tfrac{1}{2}\partial_{n}\partial_{p}g^{qr}\partial_{q}\partial_{r}g^{mp}-\tfrac{1}{2}\partial_{p}\partial_{q}\partial_{n}g^{mr}\partial_{r}g^{pq}+2A_{p}\partial_{n}g^{mp}-2g^{mp}\partial_{n}A_{p}\\
+g^{pq}\partial_{n}\partial_{p}B_{q}^{m}-g^{pq}\partial_{p}\partial_{q}B_{n}^{m}-\partial_{q}g^{mp}\partial_{n}B_{p}^{q}-\partial_{n}g^{pq}\partial_{p}B_{q}^{m}+\partial_{n}\partial_{q}g^{mp}B_{p}^{q}+2\partial_{q}g^{mp}\partial_{p}B_{n}^{q}\\
-2B_{p}^{m}B_{n}^{p}+\partial_{n}\partial_{p}g^{mq}\partial_{q}C^{p}-\partial_{p}g^{mq}\partial_{n}\partial_{q}C^{p}-2C^{p}\partial_{n}B_{p}^{m}+2\partial_{n}C^{p}B_{p}^{m}+2C^{p}\partial_{p}B_{n}^{m}=0,\label{eq:PdXeom}
\end{multline}
\end{widetext}may also be expressed as a symmetric and an antisymmetric
piece. The symmetric part is the algebraic solution for $\tilde{g}_{mn}$,
which takes the form
\begin{equation}
\tilde{g}_{mn}=g_{mn}+\mathcal{O}(\alpha'^{2}).
\end{equation}
This solution may then be replaced in $\Sigma$,
\begin{multline}
\Sigma=\tfrac{1}{(\alpha')^{2}}(g^{mn}\tilde{g}_{mn}-26)+B_{n}^{m}B_{m}^{n}+\partial_{n}C^{m}\partial_{m}C^{n}\\
+\tfrac{1}{4}\partial_{p}\partial_{q}g^{mn}\partial_{m}\partial_{n}g^{pq}-2\partial_{p}g^{mn}\partial_{m}B_{n}^{p}+12A_{m}C^{m}\\
+4\partial_{n}C^{m}B_{m}^{n}-\partial_{p}g^{mn}\partial_{m}\partial_{n}C^{p}-4g^{mn}\partial_{n}A_{m}.\label{eq:quarticHH}
\end{multline}
At leading order in $\alpha'$, the vanishing of $\Sigma$ fixes the
spacetime dimension to be 26. The next order vanishes on the support
of the other algebraic solutions.

The antisymmetric part of (\ref{eq:PdXeom}) does not depend on $\tilde{g}_{mn}$,
and can be neatly cast as
\begin{equation}
\nabla^{p}H_{mnp}=2H_{mnp}\nabla^{p}\phi.\label{eq:H-eom}
\end{equation}
The Kalb-Ramond field-strength, $H_{mnp}$, includes a gravitational Chern-Simons correction, $\Omega_{mnp}$,
\begin{eqnarray}
H_{mnp} &=& \partial_{m}b_{np}+\partial_{n}b_{pm}+\partial_{p}b_{mn}+\tfrac{\alpha'}{2}\Omega_{mnp},\label{eq:Hdef}\\
\Omega_{mnp}  &=&  \Gamma_{pr}^{q}R_{qmn}^{r}+\Gamma_{mr}^{q}R_{qnp}^{r}+\Gamma_{nr}^{q}R_{qpm}^{r} \nonumber \\ & &
 -\Gamma_{mr}^{q}\Gamma_{ns}^{r}\Gamma_{pq}^{s}+\Gamma_{mr}^{q}\Gamma_{ps}^{r}\Gamma_{nq}^{s}.\label{eq:CSterm}
\end{eqnarray}
The Christoffel symbols, $\Gamma_{mn}^{p}$, and the Riemann tensor,
$R_{pmn}^{q}$, have the usual definition,
\begin{eqnarray}
\Gamma_{mn}^{p} & = & \tfrac{1}{2}g^{pq}(\partial_{m}g_{nq}+\partial_{n}g_{mq}-\partial_{q}g_{mn}),\\
R_{pmn}^{q} & = & \partial_{m}\Gamma_{np}^{q}+\Gamma_{mr}^{q}\Gamma_{np}^{r}-(m\leftrightarrow n), 
\end{eqnarray}
with Ricci tensor $R_{mn}=R_{mpn}^{p}$ and scalar $R=g^{mn}R_{mn}$. The $\alpha'$ correction of the field-strength (\ref{eq:Hdef}) is reminiscent of Lorentz group Chern-Simons form found by Green and Schwarz \cite{Green:1984sg}. It had already been identified in the HSZ model \cite{Hohm:2014eba}.  In more practical terms, $b_{mn}$ is a
pseudo-tensor that receives $\alpha'$ corrections in the diffeomorphism
transformations. $H_{mnp}$, on the other hand, is the fully covariant
object.

Finally, $\beta_{mn}^{G}=0$ can be written as

\begin{multline}
[1-(\tfrac{\alpha'}{4}\Box)^2](R_{mn}+2\nabla_{m}\nabla_{n}\phi)= (\alpha')^{2} \mathcal{R}_{mn}  \\ 
+\tfrac{1}{4}H_{mpq}H_{n}^{\phantom{n}pq} +\tfrac{1}{4}\alpha' \nabla_o(R^{o}_{mpq}H_{n}^{\phantom{n}pq}+R^{o}_{npq}H_{m}^{\phantom{m}pq}) \\ +\alpha'(R^{p}_{mqr}H_{np}^{\phantom{np}q}+R^{p}_{nqr}H_{mp}^{\phantom{mp}q})\nabla^{r}\phi,  \label{eq:graviton-eom}
\end{multline}
with
\begin{widetext}
\vspace{-0.5cm}
\begin{multline}
\mathcal{R}_{mn}=\nabla_{m}\nabla_{p}\phi\nabla_{n}\nabla_{q}\phi\nabla^{p}\nabla^{q}\phi+\tfrac{1}{4}\nabla^{p}\phi\nabla_{p}(\nabla_{q}\phi\nabla^{q}\nabla_{m}\nabla_{n}\phi)+\tfrac{3}{2}\nabla_{p}\nabla_{q}\nabla_{m}\phi(\nabla^{p}\phi\nabla^{q}\nabla_{n}\phi-\tfrac{1}{4}\nabla^{p}\nabla^{q}\nabla_{n}\phi)\\
-\tfrac{1}{8}\Box\nabla^{p}\phi\nabla_{p}\nabla_{m}\nabla_{n}\phi-\tfrac{1}{4}\nabla^{p}(\nabla^{q}\phi\nabla_{q}\nabla_{p}\nabla_{m}\nabla_{n}\phi)-\tfrac{3}{4}\Box(\nabla_{p}\nabla_{m}\phi)\nabla^{p}\nabla_{n}\phi+\tfrac{1}{8}\nabla_{p}\phi\nabla^{p}(\nabla^{q}\phi\nabla_{q}R_{mn})\\
+\tfrac{1}{8}R_{pq}\nabla^{p}\phi\nabla^{q}\nabla_{m}\nabla_{n}\phi+\tfrac{1}{2}R_{pq}\nabla^{p}\nabla_{m}\phi\nabla^{q}\nabla_{n}\phi+\tfrac{3}{4}\nabla^{q}\phi\nabla_{q}(R_{mp}\nabla^{p}\nabla_{n}\phi)+\tfrac{1}{4}R_{mpq}^{r}\nabla^{p}\phi\nabla_{n}\nabla_{r}\nabla^{q}\phi\\
+(\tfrac{1}{4}\nabla_{r}R_{pqm}^{r}\nabla^{p}\phi+R_{mp}\nabla^{p}\nabla_{q}\phi)\nabla^{q}\nabla_{n}\phi+\tfrac{1}{8}\nabla^{q}\phi\nabla^{s}\phi R_{mpq}^{r}(5R_{rsn}^{p}-R_{nrs}^{p})+\tfrac{1}{8}\nabla_{m}R^{pq}\nabla_{n}\nabla_{p}\nabla_{q}\phi\\
-\tfrac{1}{8}\nabla_{p}\nabla_{q}R_{mn}\nabla^{p}\nabla^{q}\phi-\tfrac{3}{8}\Box R_{mp}\nabla^{p}\nabla_{n}\phi-\tfrac{1}{8}(\nabla_{p}\Box R_{mn}+R_{pq}\nabla^{q}R_{mn})\nabla^{p}\phi-\tfrac{1}{16}\nabla_{p}R_{mn}\nabla^{p}\Box\phi\\
+\tfrac{1}{8}R_{mp}[(\nabla^{q}R_{nqr}^{p}+3\nabla_{r}R_{nq}g^{pq})\nabla^{r}\phi+4R^{pq}\nabla_{q}\nabla_{n}\phi+2R_{nq}\nabla^{p}\nabla^{q}\phi-3\Box\nabla^{p}\nabla_{n}\phi]-\tfrac{1}{2}\nabla_{p}R_{mq}\nabla_{n}\nabla^{p}\nabla^{q}\phi\\
-\tfrac{1}{4}R_{mpq}^{r}[(2R_{rsn}^{p}-R_{snr}^{p})\nabla^{q}\nabla^{s}\phi+2R_{rst}^{p}\nabla^{t}\nabla_{n}\phi g^{qs}]+\tfrac{1}{8}\nabla^{s}R_{mpq}^{r}\nabla^{p}(R_{snr}^{q}-R_{rsn}^{q})+\tfrac{1}{16}\nabla_{r}R_{mp}\nabla^{r}R_{nq}g^{pq}\\
-\tfrac{3}{16}\nabla^{p}R_{mq}\nabla^{q}R_{np}-\tfrac{3}{16}R_{mp}\Box R_{nq}g^{pq}+\tfrac{1}{2}R_{qmp}^{r}(\nabla^{p}\nabla^{q}R_{nr}+\tfrac{1}{2}\nabla_{n}\nabla_{r}R^{pq})+\tfrac{1}{32}\nabla_{m}R_{pq}\nabla_{n}R^{pq}\\
+\tfrac{1}{8}R_{mp}R_{nq}R^{pq}+\tfrac{1}{8}R_{mpq}^{r}[(R_{rsn}^{p}-R_{nrs}^{p})R^{qs}+4R_{snu}^{p}R_{trv}^{q}g^{st}g^{uv}]+(m\leftrightarrow n).
\label{eq:Rmn-sixpartial}
\end{multline}
\end{widetext}
It is clear from \eqref{eq:graviton-eom} the auxiliar character of
the massive spin 2 states. They can be exactly integrated out of the
system, effectively inducing a higher derivative gravitational theory.
This is a signature feature of the bosonic and heterotic chiral strings
(see also \cite{Azevedo:2019zbn}). Their massive spectrum consists
of ghosts states with the same degrees of freedom of the first mass
level in the corresponding open string.

\section{Scattering Fields and Strings}

In conventional string theory, the background equations of motion
emerge from the Weyl anomaly in the worldsheet, order by order in
$\alpha'$. Here, on the other hand, equations (\ref{eq:dilaton-eom}),
(\ref{eq:H-eom}), and (\ref{eq:graviton-eom}) account for all $\alpha'$
corrections, completely governing the classical dynamics of the system. They can be used to determine any tree level scattering
amplitude involving the physical states (massless and massive). Alternatively,
these amplitudes may be computed directly from the chiral model \cite{LipinskiJusinskas:2019cej}
or using twisted strings \cite{Siegel:2015axg,Huang:2016bdd}. 

This interplay between (twisted) strings and fields was explored in
\cite{Guillen:2021mwp} in order to determine CFT correlators using
a purely field-theoretical input through perturbiner methods \cite{Rosly:1996vr,Rosly:1997ap,Mafra:2015gia,Mafra:2015vca,Mafra:2016ltu}.
In the present case, this input is slightly more involved because
it contains (1) gravity, and (2) a higher derivative kinetic term
due to the spin 2 ghosts. Fortunately, the perturbiner was extended
to gravitational theories coupled to matter in \cite{Gomez:2021shh} (and later in \cite{Cho:2021nim}, in the context of doube field theory).
The higher derivative kinetic term is but a small detour of the traditional
(quadratic) case.

A necessary step in working with the perturbiner is to fix the gauge
symmetries of the field theory. We will work here with\begin{subequations}\label{eq:gauge}
\begin{eqnarray}
\eta^{np}\partial_{p}b_{mn} & = & 0,\\
\eta^{np}\Gamma_{np}^{m}+2\nabla^{m}\phi & = & 0,\label{eq:dDdgauge}
\end{eqnarray}
\end{subequations}fixing the form symmetry and the spacetime
diffeomorphisms. The latter, which we dub dilaton-de Donder gauge,
is a convenient choice for our \emph{string frame} equations of motion.
As opposed to the Einstein frame, with metric $g_{mn}^{(E)}=e^{-\phi/6}g_{mn}$,
the string frame intermingles the dilaton and the graviton. Indeed,
the linearized solutions of (\ref{eq:dilaton-eom}), (\ref{eq:H-eom}),
and (\ref{eq:graviton-eom}), i.e. single-particle states, can be
parametrized as\begin{subequations}
\begin{eqnarray}
\bar{b}_{mn}(x) & = & b_{mn}e^{ik^{b}\cdot x},\\
\bar{g}_{mn}(x) & = & \eta_{mn}+\tfrac{1}{6}\eta_{mn}\phi e^{ik^{\phi}\cdot x}\nonumber \\
 &  & +g_{mn}^{\pm}e^{ik^{\pm}\cdot x}+g_{mn}^{0}e^{ik^{g}\cdot x},\\
\bar{\phi}(x) & = & \phi e^{ik^{\phi}\cdot x},
\end{eqnarray}
\end{subequations}where $(k^{\pm})^{2}=\mp4/\alpha'$, and $k^{2}=0$
otherwise. The polarizations denote the dilaton, $\phi$, the Kalb-Ramond
2-form, $b_{mn}$, the graviton, $g_{mn}^{0}$, and the ghosts, $g_{mn}^{\pm}$.
They are traceless and transversal with the respective momenta, and
there is a residual gauge symmetry $\delta b_{mn}=k_{m}\gamma_{n}-k_{n}\gamma_{m}$,
and $\delta g_{mn}^{0}=k_{m}\lambda_{n}+k_{n}\lambda_{m}$, with $k\cdot\gamma \equiv \eta^{mn} k_m \gamma_n =0$.

Next, we define the multiparticle expansions:\begin{subequations}\label{eq:multiparticle}
\begin{eqnarray}
b_{mn}(x) & = & \sum_{J}B_{Jmn}e^{ik_{J}\cdot x},\\
g_{mn}(x) & = & \eta_{mn}+\sum_{J}G_{Jmn}e^{ik_{J}\cdot x},\\
\phi(x) & = & \sum_{J}\Phi_{J}e^{ik_{J}\cdot x}.
\end{eqnarray}
\end{subequations}The sums are over all ordered
words $J=j_{1}j_{2}\ldots j_{n}$ ($j_{1}<...<j_{n}$) composed of
single-particle labels $j$ (letters). The multiparticle momenta are
given by $k_{J}=k_{j_{1}}+...+k_{j_{n}}$. Multiparticle currents
are denoted by $B_{Jmn}$, $G_{Jmn}$, and $\Phi_{J}$, and reduce
to single-particle polarizations for one-letter words $J=j$. See \cite{Gomez:2021shh} for notation.

In order to satisfy the classical equations of motion, the expansions
(\ref{eq:multiparticle}) lead to recursive definitions of the multiparticle
currents of the form\begin{subequations}
\begin{eqnarray}
s_{J}B_{Jmn} & = & B_{Jmn}^{\textrm{int}},\\
( 1+\tfrac{\alpha'}{4}s_{J}) ( 1-\tfrac{\alpha'}{4}s_{J})  s_{J}G_{Jmn} & = & G_{Jmn}^{\textrm{int}},\\
s_{J}(\Phi_{J}-\tfrac{1}{6}\eta^{mn}G_{Jmn}) & = & \Phi_{J}^{\textrm{int}},
\end{eqnarray}
\end{subequations}with $s_{J}=k_{J}^{2}$ denoting the generalized
Mandelstam variables. The interaction terms on the right hand side
are built out of powers of multiparticle currents with subwords of
$J$. For instance,
\begin{equation}
\Phi_{J}^{\textrm{int}}=2\sum_{J=K\cup L} (k_{K} \cdot k_{L}) \Phi_{K}\Phi_{L}+\ldots,
\end{equation}
where the sum is over all possible ways of partitioning the ordered word $J$ into two ordered subwords $K$
and $L$, such that $K\cup L=J$. The ellipsis also involve splittings with more than two subwords.
The precise forms of $\{B_{Jmn}^{\textrm{int}},G_{Jmn}^{\textrm{int}},\Phi_{J}^{\textrm{int}}\}$ are lengthy but straightforward to determine via the
equations of motion. More details will be given in \cite{NRS-long}.

We are now ready to compute the tree level scattering amplitudes of
the model, the final step in the perturbiner method. They are defined
as\begin{subequations}\label{eq:amplitudes-perturbiner}
\begin{eqnarray}
\mathcal{A}_{b} & = & b^{mn}_{1}B^{\textrm{int}}_{2...Nmn}, \label{eq:amp-b}\\
\mathcal{A}_{g} & = &(g^{0mn}_{1}+g^{\pm mn}_{1})G^{\textrm{int}}_{2...Nmn}, \label{eq:amp-g}\\
\mathcal{A}_{\phi} & = & \tfrac{2}{3} \phi_{1}(\Phi^{\textrm{int}}_{2...N}+\tfrac{1}{6} \eta^{mn} G^{\textrm{int}}_{2...Nmn}), \label{eq:amp-phi}
\end{eqnarray}
\end{subequations}
on the support of momentum conservation, $k_{1...N}=0$. For example, the three-point amplitude involving the dilaton and two Kalb-Ramond forms, using either \eqref{eq:amp-b} or \eqref{eq:amp-phi}, is given by
\begin{equation}
\mathcal{A}(b_1,b_2,\phi_3) = \tfrac{1}{6}[(k_{2}-k_{3})\cdot b_{1}\cdot b_{2}\cdot(k_{3}-k_{1})]\phi_{3}. \label{eq:3ptbbphi}
\end{equation}

The residual
gauge symmetries of the single-particle polarizations $b_{mn}$ and
$g_{mn}^{0}$ leave the amplitudes invariant in the gauge (\ref{eq:gauge}),
which translates to multiparticle currents as\begin{subequations}\label{eq:gauge-currents}
\begin{eqnarray}
k_{J}^{n}B_{Jmn} & = & 0,\\
k_{J}^{n}G_{Jmn} & = & k_{Jm}(\tfrac{1}{2}\eta^{np}G_{Jnp}-2\Phi_{J}).
\end{eqnarray}
\end{subequations}

As the ultimate consistency check of the field equations (\ref{eq:dilaton-eom}),
(\ref{eq:H-eom}), and (\ref{eq:graviton-eom}), we should compare
the amplitudes (\ref{eq:amplitudes-perturbiner}) with the ones computed
using the chiral string CFT. There, the unintegrated vertex operators describing the physical states take
the form
\begin{eqnarray}
U_{0} & = & c\tilde{c}P_{+}^{m}P_{-}^{n}(b_{mn}+g_{mn}^{0}+\tfrac{1}{6}\hat{\eta}_{mn}\phi)e^{ik\cdot X},\\
U_{\pm} & = & c\tilde{c}P_{\pm}^{m}P_{\pm}^{n}g_{mn}^{\pm}e^{ik\cdot X},
\end{eqnarray}
where $P_{\pm}^{m}=(P^{m}\pm\tfrac{1}{\alpha'}\partial X^{m})$. In the dilaton vertex, $\hat{\eta}_{mn}=\eta_{mn}-k_{m}q_{n}-k_{n}q_{m}$, where $q_m$ is a reference vector satisfying $q\cdot k=1$ and $q^2=0$.  Now, using
the results of \cite{LipinskiJusinskas:2019cej}, we were able to
cross check all three-point and an assortment of four-point tree level
amplitudes of the bosonic chiral string (see also \cite{Huang:2016bdd,Leite:2016fno}) and the output described above. For example, the chiral string analogous of \eqref{eq:3ptbbphi} is given by
\begin{equation}
\mathcal{A}^{\textrm{string}}(b_1,b_2,\phi_3) = \mathcal{N}\left(\tfrac{2}{\alpha'}\right)^{2}\mathcal{A}(b_1,b_2,\phi_3),
\end{equation}
where $\mathcal{N}$ depends on the CFT data and vertex normalizations. An overall numerical factor cannot be fixed using the perturbiner method, so this is the only extra input from the chiral string. Once fixed, all tree level amplitudes match exactly. 

\section{Summary \& Prospects}

In this work we have derived the \emph{$\alpha'$-exact} background
field equations of the HSZ model (bosonic chiral string), i.e. equations \eqref{eq:dilaton-eom}, \eqref{eq:H-eom}, and \eqref{eq:graviton-eom}. Up to our knowledge, this is the first example
of a string model in which all $\alpha'$ corrections have been completely
determined for a generic background.

In order to support our results, we performed a thorough cross check
between tree level amplitudes computed using two fundamentally different
approaches. Namely, the usual string prescription and the field-theoretical
perturbiner method. Furthermore, the background field equations have the expected
$\alpha'\to\infty$ and $\alpha'\to0$ limits. The former is the
bosonic ambitwistor string, corresponding to a fully covariantized
version of \cite{Berkovits:2018jvm}. The latter is given by the conventional
bosonic string (see e.g. \cite{Polchinski:1998rq}). We note, in particular,
that the $\alpha'\to0$ limit is obscured in the worldsheet action
and the BRST charge. This suggests that a different set of variables
might be more suitable to effectively built a worldsheet theory for (super) gravity. For finite $\alpha'$, two states emerge, with opposite mass squared. Indeed, the model is $\alpha'\to-\alpha'$ symmetric. 
Unlike in massive bigravity \cite{deRham:2010kj,Hassan:2011hr}, these spin 2 states have
an auxiliary role and can be integrated out of the equations of motion,
effectively introducing a higher derivative kinetic term with ghosts (see also \cite{Lust:2023sfk} and references therein for other results on massive spin 2 fields in string theory). A separate paper is being prepared with an extended, more detailed presentation of the results introduced here \cite{NRS-long}. It would still be interesting to investigate whether equations
\eqref{eq:dilaton-eom}, \eqref{eq:H-eom}, and \eqref{eq:graviton-eom}  have nontrivial solutions (such as ghost condensates \cite{Arkani-Hamed:2003pdi}, also suggested in \cite{Hohm:2016lim}) and, if so, their
classical properties. Additionally, one could also investigate if and how T-duality is manifested in these equations  (see also \cite{Hronek:2020xxi} for a related discussion). In the chiral formulation, it is not clear how to identify, for instance, the winding modes of $X^m$, though this can be easily done in the second order formulation \cite{Casali:2017mss}.

The HSZ model was considered the worldsheet theory behind the so-called \textit{doubled $\alpha'$ geometry} \cite{Hohm:2013jaa}, which has been ambitiously developed in several directions (to cite a few \cite{Marques:2015vua,Hohm:2019ccp,Hohm:2019jgu,Codina:2022onm}). Our results directly clash with different ideas related to double $\alpha'$ theories. Spectrum wise, and as expected \cite{Siegel:2015axg,Huang:2016bdd,Azevedo:2019zbn}, we observe the usual massless spectrum of the closed string plus two massive spin two states. This is confirmed by the analysis of physical poles of the amplitudes. There are no massive scalars, unlike in \cite{Hohm:2016lim}. Relatedly, the massive fields can be fully integrated out and lead to a finite series of higher-derivative corrections (i.e. a \textit{finite} number of $\alpha'$ corrections). The background field equations we obtain have no $(\alpha')^3$ corrections or higher. Barring some intriguing similarities, such as Green-Schwarz like corrections to the Kalb-Ramond field strength, and involving up to six derivatives, the doubled $\alpha'$ geometry and the bosonic chiral string lead to different spacetime theories. Therefore, unless further constraints are imposed, the HSZ model cannot be seen as a consistent truncation of string theory.

Besides providing a more complete picture of the bosonic chiral string, we can mimic the construction of \cite{Guillen:2021mwp} in order to convert field theory amplitudes \eqref{eq:amplitudes-perturbiner} to their respective chiral string amplitudes \textit{before} the moduli space integrations. The reasoning is more or less straightforward. First, we have to remember that the chiral worldsheet is equivalent to a second order worldsheet with target space coordinates satisfying \cite{Siegel:2015axg}
\begin{equation}
X^{m}(z,\bar{z})X^{n}(y,\bar{y})\sim\tfrac{\alpha'}{2}\eta^{mn}[\ln(\bar{z}-\bar{y})-\ln(z-y)],
\end{equation}
with a sign flip between the holomorphic and antiholomorphic sectors. The CFT correlators factorize into a chiral and an antichiral piece, each matching a given open-string-like configuration involving massless vectors and massive spin 2 states. Since the chiral correlators are universal, the amplitudes computed via \eqref{eq:amplitudes-perturbiner} can be effectively translated to ordinary open and closed string scattering at tree level. The method is analogous to the one described in \cite{Guillen:2021mwp}, but now involving a colorless theory. The bosonic case is particularly interesting, since the relevant CFT correlators involving massive states are more tedious to compute. It should, nonetheless, be extended to the more interesting ones of the spinning string. They involve additional ingredients such as picture changing, spin fields, and we plan to address this in a future work.

All in all, the interplay between fields and strings has been a prolific research direction, with several exciting developments (see e.g. \cite{Azevedo:2018dgo,Adamo:2013tsa,He:2017spx,Geyer:2015bja,Cachazo:2015aol,Geyer:2016wjx,Geyer:2018xwu,Mizera:2017rqa,He:2018pol,Geyer:2021oox} and related works). While most of their impact concerns massless theories, we expect our outcome will help to understand their extension to massive theories, including higher spin fields (cf. \cite{Jusinskas:2021bdj}), at tree and loop level.

\begin{acknowledgments}
We thank T. Azevedo, S. Chakrabarti, T. Erler, P. R. S. Gomes, O. Schlotterer for useful discussions, comments on the draft, and reference suggestions. We also thank Chrysoula Markou for related discussions. Part of our symbolic computations were performed with the support of Cadabra \cite{cadabra1,cadabra2,cadabra3}. RLJ acknowledges the financial support from the Czech Academy of Sciences under the project number LQ100102101.
\end{acknowledgments}

\end{document}